\documentclass{ws-ijmpd}

\begin{document}

\markboth{Slava G. Turyshev and Michael Shao}
{The Laser Astrometric Test of Relativity}

\catchline{}{}{}{}{}
%
%%%%%%%%%%%%%%%%%%%%%%%%%%%%%%%%%%%%%%%%%%%%%%%%%%%%%%%%%%%%%%%%%%%%

%\title{The Laser Astrometric Test of Relativity: \\ Science, Technology, and Mission Design}

\title{LASER ASTROMETRIC TEST OF RELATIVITY: \\SCIENCE, TECHNOLOGY, AND MISSION DESIGN}

%\author{Slava G.\ Turyshev and Michael Shao}
\author{SLAVA G. TURYSHEV and MICHAEL SHAO}

\address{Jet Propulsion Laboratory, 
California Institute of Technology, \\
4800 Oak Grove Drive, Pasadena, CA 91109, USA\\
turyshev@jpl.nasa.gov, mshao@huey.jpl.nasa.gov}

%\author{SECOND AUTHOR}
%
%\address{Group, Laboratory, Address\\
%City, State ZIP/Zone, Country\\
%second\_author@group.com}

\maketitle

%\begin{history}
%\received{11 August 2006}
%%\revised{Day Month Year}
%\comby{Slava G. Turyshev}
%\end{history}

\begin{abstract}
The Laser Astrometric Test of Relativity (LATOR) experiment is designed to explore general theory of relativity in the close proximity to the Sun -- the most intense gravitational environment in the solar system. Using independent time-series of highly accurate measurements of the Shapiro time-delay (interplanetary laser ranging accurate to 3~mm at $\sim$2AU) and interferometric astrometry (accurate to 0.01 picoradian), LATOR will measure gravitational deflection of light by the solar gravity with accuracy of 1 part in a billion -- a factor $\sim$30,000 better than currently available.  LATOR will perform series of highly-accurate tests in its search for cosmological remnants of scalar field in the solar system.   We present science, technology and mission design for the LATOR mission.
\end{abstract}


%*********************1) INTRODUCTION
\section{Motivation}
%\label{sec:intro}
\label{sec:sci_mot}

Recent remarkable progress in observational cosmology has again submitted general theory of  relativity to a test by suggesting a non-Einsteinian model of universe's evolution. From the theoretical standpoint, the challenge is even stronger---if the gravitational field is to be quantized,  general relativity will have to be modified. Additionally, recent advances in the scalar-tensor extensions of gravity have intensified searches for very small deviations from the Einstein's theory, at the level of three to five orders of magnitude below the level currently tested by experiment. 

The Laser Astrometric Test of Relativity (LATOR) is a proposed space-based experiment to significantly improve the tests of relativistic gravity\cite{lator_cqg_2004,lator_LCDF-2006}. LATOR is designed to address the questions of fundamental importance by searching for a cosmologically-evolved scalar field that is predicted by  modern theories, notably the string theory. LATOR will also test theories that attempt to explain the small acceleration rate of the Universe (so-called `dark energy') via modification of gravity at very large distances, notably brane-world theories. 

Section \ref{sec:sci_mot} of this paper discusses the theoretical framework, science motivation and objectives for LATOR. Section \ref{sec:lator_description} provides an overview of mission and optical designs for LATOR. Section~\ref{sec:conc} discusses the next steps in  mission development.  

\subsection{The PPN Formalism}

Generalizing on the Eddington's phenomenological parameterization of the gravitational metric tensor field, a method called the parameterized post-Newtonian (PPN) formalism has been developed\cite{Ken_EqPrinciple68b,Will_book93}.
This method  represents the gravity tensor's potentials for slowly moving bodies and weak interbody gravity, and it is valid for a broad class of metric theories including general relativity as a unique case.  The several parameters in the PPN metric expansion vary from theory to theory, and they are individually associated with various symmetries and invariance properties of underlying theory.  An ensemble of experiments can be analyzed using the PPN metric, to determine the unique value for these parameters, and hence the metric field, itself.
In locally Lorentz-invariant theories the expansion of the metric field for a single, slowly-rotating gravitational source in PPN coordinates\cite{Ken_EqPrinciple68b,Will_book93} is given by:
\begin{eqnarray}
\label{eq:metric}
g_{00} &=& 1-2\frac{M}{r}Q(r,\theta) +2\beta\frac{M^2}{r^2}+{\cal O}(c^{-6}),\nonumber\\ 
g_{0i}  &=&  2(\gamma+1)\frac{[\vec{J}\times \vec{r}]_i}{r^3}+
{\cal O}(c^{-5}),\\ 
g_{ij}  &=&  -~\delta_{ij}\Big[1+
2\gamma \frac{M}{r} Q(r,\theta)+
\frac{3}{2}\delta \frac{M^2}{r^2}\Big]
\hskip -1pt +\hskip -0pt {\cal O}(c^{-6}),\nonumber
\end{eqnarray}

\noindent where $M$ and $\vec J$ being the mass and angular momentum of the Sun, $Q(r,\theta)=1-J_2\frac{R^2}{r^2}\frac{3\cos^2\theta-1}{2}$, with $J_2$ being the quadrupole moment of the Sun and $R$ being its radius.  $r$ is the distance between the observer and the center of the Sun.  $\beta, \gamma, \delta$ are the PPN parameters (in general relativity $\beta=\gamma=\delta =1$). The $M/r$ term in $g_{00}$ component is the Newtonian limit; the terms multiplied by the post-Newtonian parameters $\beta, \gamma$,  are post-Newtonian terms. The term multiplied by the post-post-Newtonian parameter  $\delta$ also enters the calculation of the relativistic light deflection\cite{Ken_cqg96}.

The most precise value for the PPN parameter  $\gamma$ is at present given by the Cassini mission\cite{cassini_ber} as: $\gamma -1 = (2.1\pm2.3)\times10^{-5}$. Using the recent Cassini result\cite{cassini_ber} on the PPN  $\gamma$, the parameter $\beta$ was measured as $\beta-1=(0.9\pm1.1)\times 10^{-4}$ from lunar laser ranging (LLR)\cite{LLR_beta_2004}. The PPN parameter $\delta$ has not yet been measured though its value can be inferred from other measurements.

The Eddington parameter $\gamma$, whose value in general relativity is unity, is perhaps the most fundamental PPN parameter, in that $\frac{1}{2}(1-\gamma)$ is a measure, for example, of the fractional strength of the scalar gravity interaction in scalar-tensor theories of gravity\cite{Damour_EFarese96a}.  Within perturbation theory for such theories, all other PPN parameters to all relativistic orders collapse to their general relativistic values in proportion to $\frac{1}{2}(1-\gamma)$. Thus, a measurement of the first order light deflection effect at the level of accuracy comparable with the second-order contribution would provide the crucial information separating alternative scalar-tensor theories of gravity from general relativity\cite{Ken_EqPrinciple68b} and also to probe possible ways for gravity quantization and to test modern theories of cosmological evolution\cite{Damour_Nordtvedt_93a}\cdash\cite{DPV02a} (see Sec.~\ref{sec:mot}). The LATOR mission is designed to directly address this issue with an unprecedented accuracy.

\subsection{Motivations for Precision Gravity Experiments}
\label{sec:mot} 

%\subsubsection{Tensor-Scalar Theories of Gravity}
%\label{sec:mot_theories}

Recent theoretical findings suggest that the present agreement between general relativity and experiment might be naturally compatible with the existence of a scalar contribution to gravity. Damour and Nordtvedt in Ref.~\refcite{Damour_Nordtvedt_93a}\footnote{See also Ref.~\refcite{DamourPolyakov94a} for non-metric versions of this mechanism together with Ref.~\refcite{DPV02a} for the recent summary of a dilaton-runaway scenario.} have found that a scalar-tensor theory of gravity may contain a ``built-in'' cosmological attractor mechanism toward general relativity.  These scenarios assume that the scalar coupling parameter $\frac{1}{2}(1-\gamma)$ was of order one in the early universe, and show that it then evolves to be close to, but not exactly equal to, zero at the present time\cite{lator_cqg_2004}. 
Ref.~\refcite{Damour_Nordtvedt_93a} estimates the likely order of magnitude of the left-over coupling strength at present time which, depending on the total mass density of the universe, can be given as $1-\gamma \sim 7.3 \times 10^{-7}(H_0/\Omega_0^3)^{1/2}$, where $\Omega_0$ is the ratio of the current density to the closure density and $H_0$ is the Hubble constant in units of 100 km/sec/Mpc. Compared to the cosmological observations a lower bound of $(1-\gamma) \sim 10^{-6}-10^{-7}$ can be derived for the present value of PPN parameter $\gamma$. 

Ref.~\refcite{DPV02a} estimated $\frac{1}{2}(1-\gamma)$ within the framework compatible with string theory and modern cosmology confirming the previous result\cite{Damour_Nordtvedt_93a}. This analysis discusses a scenario when a composition-independent coupling of dilaton to hadronic matter produces detectable deviations from general relativity in high-accuracy light deflection experiments in the solar system. This work assumes only some general properties of the coupling functions and then only assumes that $(1-\gamma)$ is of order of one at the beginning of the controllably classical part of inflation.
It was shown\cite{DPV02a} that one can relate the present value of $\frac{1}{2}(1-\gamma)$ to the cosmological density fluctuations.
For the simplest inflationary potentials favored by WMAP mission\cite{[4c]} (i.e. $m^2 \chi^2$) found that the present value of $(1-\gamma)$ could be just below $10^{-7}$.
In particular, within this framework the value of $\frac{1}{2}(1-\gamma)$ depends on the model taken for the inflation potential; thus for  $V(\chi)\propto\chi^2$, with $\chi$ being the inflation field, the level of the expected deviations from general relativity is $\sim0.5\times10^{-7}$ (see Ref.~\refcite{DPV02a}). Note that these predictions are based on the work in scalar-tensor extensions of gravity which are consistent with, and indeed often part of, present cosmological models. 

The analyses above motivate new searches for small deviations of relativistic gravity in the solar system by predicting that such deviations are currently present in the range from $10^{-5}$ to $5\times10^{-8}$ for $\frac{1}{2}(1-\gamma)$, i.e. for observable post-Newtonian deviations from general relativity predictions and, thus, should be easily detectable with LATOR. This would require measurement of the effects of the next post-Newtonian order ($\propto G^2$) of light deflection resulting from gravity's intrinsic non-linearity. An ability to measure the first order light deflection term at the accuracy comparable with the effects of the second order is of the utmost importance and a major challenge for the 21$^{\rm st}$ century fundamental physics.  

%\subsubsection{Experimental Motivations for High-Accuracy Gravity Tests}
%\label{sec:mot_experiment}

There is now multiple evidence indicating that 70\% of the critical density of the universe is in the form of a ``negative-pressure'' dark energy component; there is no understanding as to its origin and nature. The fact that the expansion of the universe is currently undergoing a period of acceleration now seems rather well tested: it is directly measured from the light-curves of several hundred type Ia supernovae\cite{perlmutter99}\cdash\cite{[3c]}, and independently inferred from observations of CMB by WMAP mission\cite{[4c]} and other CMB experiments\cite{[6c],[5c]}. Cosmic speed-up can be accommodated within general relativity by invoking a mysterious cosmic fluid with large negative pressure, dubbed dark energy. The simplest possibility for dark energy is a cosmological constant; unfortunately, the smallest estimates for its value are 55 orders of magnitude too large. Most of the theoretical studies operate in the shadow of the cosmological constant problem, the most embarrassing hierarchy problem in physics. This fact has motivated other possibilities, most of which assume $\Lambda=0$, with the dynamical dark energy being associated with a new scalar field. However, none of these suggestions is compelling and most have serious drawbacks. 

Given the challenge of this problem, a number of authors considered the possibility that cosmic acceleration is not due to some kind of stuff, but rather arises from new gravitational physics (see discussion in Refs.~\refcite{PeeblesRatra03,[carroll],Carroll_HT_03}). In particular, some extensions to general relativity in a low energy regime\cite{[carroll]} were shown to predict an experimentally consistent evolution of the universe without the need for dark energy. These models are expected to produce measurable contribution to the parameter $\gamma$  in experiments conducted in the solar system also at the level of $1-\gamma \sim 10^{-7}-5\times10^{-9}$, thus further motivating the relativistic gravity research. Therefore, the PPN parameter $\gamma$ may be the only key parameter that holds the answer to most of the questions discussed above.\footnote{Also an anomalous parameter $\delta$ will most likely be accompanied by a `$\gamma$ mass' of the Sun which differs from the gravitational mass of the Sun and therefore will show up as anomalous $\gamma$.}

In summary, there are a number of theoretical and experimental reasons to question the validity of general relativity; LATOR will address theses challenges.
%We shall now discuss the LATOR mission in more details.

\section{Overview of LATOR}
\label{sec:lator_description}

The LATOR experiment uses the standard technique of time-of-fight laser ranging between two micro-spacecraft whose lines of sight pass close by the Sun and also a long-baseline stellar optical interferometer (placed above the Earth's atmosphere) to accurately measure deflection of light by the solar gravitational field in the extreme proximity to the Sun\cite{lator_cqg_2004}.  Figure \ref{fig:lator} shows the general concept for the LATOR missions including the mission-related geometry, experiment details  and required accuracies.  

\subsection{Mission Design and Anticipated Performance}
\label{sec:lator_science}

LATOR is a Michelson-Morley-type experiment designed to test the pure tensor metric nature of gravitation -- a fundamental postulate of Einstein's theory of general relativity\cite{lator_cqg_2004,lator_LCDF-2006}.  With its focus on gravity's action on light propagation it complements other tests which rely on the gravitational dynamics of bodies.  LATOR relies on a combination of independent time-series of highly accurate measurements of the gravitational deflection of light in the immediate proximity to the Sun along with measurements of the Shapiro time delay on the interplanetary scales (to a precision respectively better than $0.01$~prad and 3 mm).  Such a combination of observables is unique and enables LATOR to significantly improve tests of relativistic gravity.

%************
\begin{figure*}[t!]
 \begin{center}
\noindent  
\psfig{figure=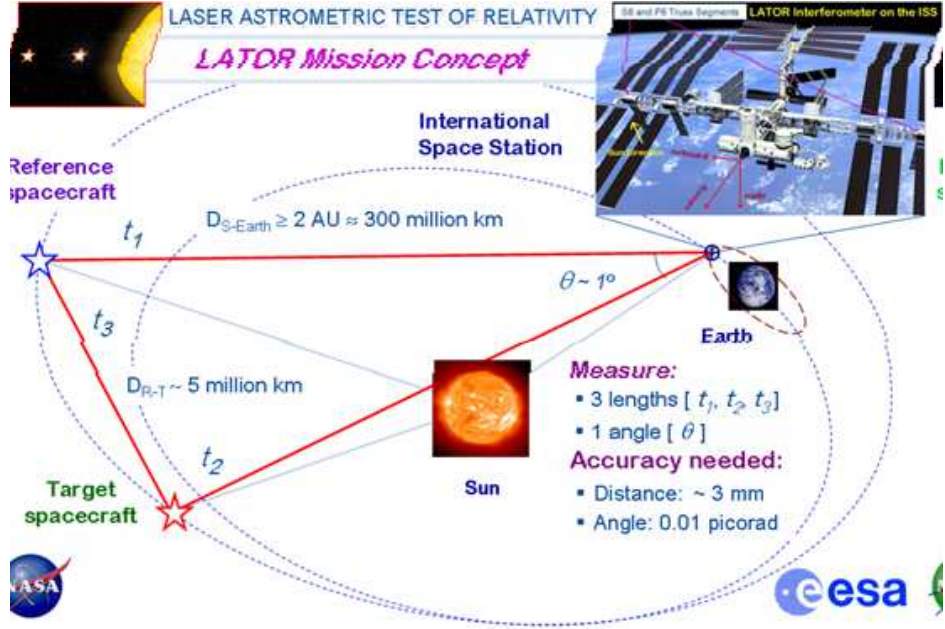,width=124mm}
\end{center}
\vskip -10pt 
  \caption{The overall geometry of the LATOR experiment.} 
\label{fig:lator}
\vskip -5pt 
\end{figure*} 
%**************

The schematic of the LATOR experiment is given in Figure~\ref{fig:lator}. Two spacecraft are injected into a heliocentric solar orbit on the opposite side of the Sun from the Earth. The triangle in the figure has three independent quantities but three arms are monitored with laser metrology. Each spacecraft equipped with a laser ranging system that enables a measurement of the arms of the triangle formed by the two spacecraft and the ISS.   According to Euclidean rules this determines a specific angle at the interferometer; LATOR can measure this angle directly. In particular, the laser beams transmitted by each spacecraft are detected by a long baseline ($\sim$ 100 m) optical interferometer on the ISS. The actual angle measured at the interferometer is compared to the angle calculated using Euclidean rules and three side measurements; the difference is the non-Euclidean deflection signal (which varies in time during spacecraft passages), which contains the scientific information. This built-in redundant--geometry optical truss eliminates the need for drag-free spacecraft for high-accuracy navigation\cite{lator_cqg_2004}. 

The uniqueness of LATOR comes with its geometrically redundant architecture that enables it to measure the departure from Euclidean geometry ($\sim 8\times 10^{-6}$ rad) caused by the solar gravity field, to a very high accuracy.   This departure is shown as a difference between the calculated Euclidean value for an angle in the triangle and its value directly measured by the interferometer.  This discrepancy, which results from the curvature of the space-time around the Sun and can be computed for every alternative theory of gravity, constitutes LATOR's signal of interest. The precise  measurement of this departure constitutes the primary mission objective.

The LATOR's primary mission objective is to measure the key PPN parameter $\gamma$ with an accuracy of a part in 10$^9$.  
When the light deflection in solar gravity is concerned, the magnitude of the first order effect as predicted by general relativity for the light ray just grazing the limb of the Sun is $\sim1.75$ arcsec. The effect varies inversely with the impact parameter. The second order term is almost six orders of magnitude smaller resulting in  $\sim 3.5$ microarcsec ($\mu$as) light deflection effect, and which falls off inversely as the square of the light ray's impact parameter\cite{lator_cqg_2004,epstein_shapiro_80}\cdash\cite{RichterMatzner82a}. The relativistic frame-dragging term
is $\pm 0.7 ~\mu$as, and contribution of the solar quadrupole moment, $J_2$, is sized as 0.2 $\mu$as (using theoretical value of the solar quadrupole moment $J_2\simeq10^{-7}$). The small magnitudes of the effects emphasize the fact that, among the four forces of nature, gravitation is the weakest interaction; it acts at very long distances and controls the large-scale structure of the universe, thus, making the precision tests of gravity a very challenging task. 

The first order effect of light deflection in the solar gravity caused by the solar mass monopole is $\alpha_1=1.75$ arcsec, which corresponds to an interferometric delay of $d\simeq b\alpha_1\approx0.85$~mm on a $b=100$~m baseline. Using laser interferometry, we currently are able to measure distances with an accuracy (not just precision but accuracy) of $\leq$~1~pm. In principle, the 0.85 mm gravitational delay can be measured with $10^{-10}$ accuracy versus $10^{-4}$ available with current techniques. However, we use a conservative estimate for the delay of 5 pm which would produce the measurement of $\gamma$ to accuracy of 1 part in $10^{9}$ (rather than 1 part in $10^{-10}$), which would be already a factor of 30,000 accuracy improvement when compared to the recent Cassini result\cite{cassini_ber}. Furthermore, we have targeted an overall measurement accuracy of 5 pm per measurement, which for $b=100$~m this translates to the accuracy of 0.05 prad $\simeq0.01~\mu$as. With 4 measurements per observation, this yields an accuracy of $\sim5.8\times 10^{-9}$ for the first order term. The second order light deflection is approximately 1700 pm and, with 5 pm accuracy and $\sim~400$ independent data points, it could be measured with accuracy of $\sim1$ part in $10^{4}$, including first ever measurement of the PPN parameter $\delta$.  The frame dragging effect would be measured with $\sim 1$ part in $10^{3}$ accuracy and the solar quadrupole moment can be modestly measured to 1 part in 200, all with respectable signal to noise ratios.  Recent covariance studies performed for the LATOR mission\cite{hellings_2005,Ken_lator05} confirm the design performance parameters and also provide valuable recommendations for further mission developments. 

We shall now consider the LATOR mission architecture.

\subsection{Mission Architecture: Evolving Light Triangle}
\label{sec:triangle}

The LATOR mission architecture uses an evolving light triangle formed by laser ranging between two spacecraft (placed in $\sim$1 AU heliocentric orbits) and a laser transceiver terminal on the International Space Station (ISS), via European collaboration.  The objective is to measure the gravitational deflection of laser light as it passes in extreme proximity to the Sun (see Figure \ref{fig:lator}).  To that extent, the long-baseline ($\sim$100 m) fiber-coupled optical interferometer on the ISS will perform differential astrometric measurements of the laser light sources on the two spacecraft as their lines-of-sight pass behind the Sun.  As seen from the Earth, the two spacecraft will be separated by about 1$^\circ$, which will be accomplished by a small maneuver immediately after their launch\cite{lator_cqg_2004,stanford_texas}. This separation would permit differential astrometric observations to an accuracy of $\sim 10^{-13}$ radians needed to significantly improve measurements of gravitational deflection of light in the solar gravity.

%
%
%\subsection{Spacecraft Trajectory: a 3:2 Earth Resonant Orbit}
%\label{sec:3:2orbit}

\begin{figure*}[t!]
% \begin{center}
\hskip -10pt 
\begin{minipage}[b]{.46\linewidth}
\centering \psfig{figure=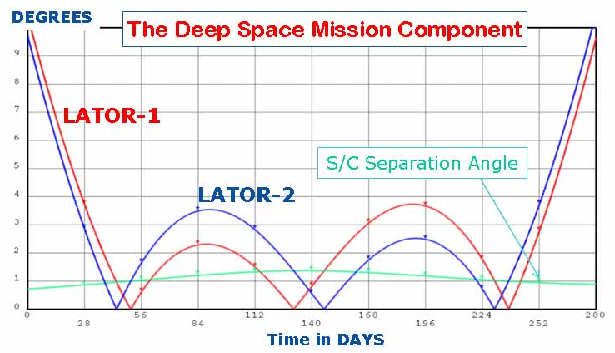,width=82mm}
\end{minipage}
\hfill  
\begin{minipage}[b]{.32\linewidth}
\centering 
\vbox{\psfig{figure=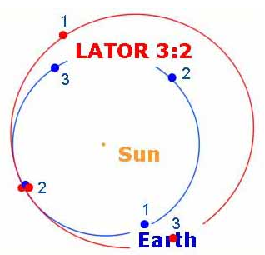,width=3.2cm}\\\vskip26pt}
\end{minipage}
\caption{Left: The Sun-Earth-Probe angle during the period of 3 occultations (two periodic curves) and the angular separation of the spacecraft as seen from the Earth (lower smooth line). Time shown is days from the moment when one of the spacecraft are at 10º distance from the Sun. Right: View from the North Ecliptic of the LATOR spacecraft in a 3:2 resonance. The epoch is taken near the first occultation.  
 \label{fig:lator_sep_angle2}}
% \end{center}
\vskip -5pt 
%\end{figure*}
%
%%*********************************
%%************
%\begin{figure*}[t!]
 \begin{center}
\noindent    
\psfig{figure=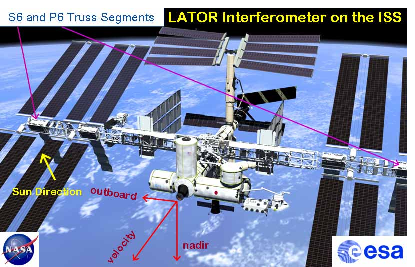,width=95mm}%,height=90mm}
%\end{center}
%\vskip -10pt 
%  \caption{Text.  
% \label{fig:lator_sep_angle}}
%\end{figure*} 
%
%%**************
%%************
%\begin{figure}[t!]
% \begin{center}
%\noindent    
\psfig{figure=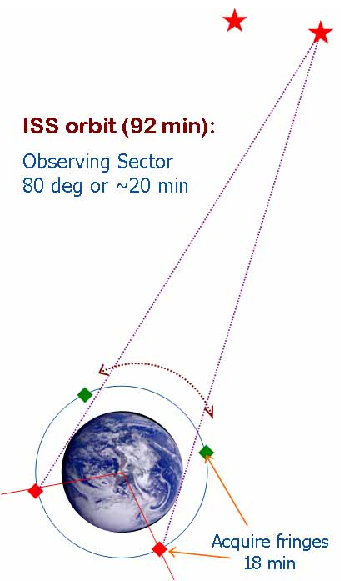,width=3.4cm}%,height=90mm}
%\end{center}
%\vskip -10pt 
  \caption{Left: Location of the LATOR interferometer on the ISS. To utilize the inherent ISS Sun-tracking capability, the LATOR optical packages will be located on the outboard truss segments P6 and S6 outwards. Right: Signal acquisition for each orbit of the ISS; variable baseline allows for solving fringe ambiguity.  
 \label{fig:iss_config}}
 \end{center}
\end{figure*}

To enable the primary objective, LATOR will place two spacecraft into a heliocentric orbit, to provide conditions for observing the spacecraft when they are behind the Sun as viewed from the ISS (see Figures~\ref{fig:lator_sep_angle2},\ref{fig:iss_config}).  
Figure \ref{fig:lator_sep_angle2} shows the  trajectory  and the occultations in more details.  The figure on the right is the spacecraft position in the solar system showing the Earth's and LATOR's orbits relative to the Sun.  The epoch of this figure shows the spacecraft passing behind the Sun as viewed from the Earth.  The  figure on the left shows the trajectory when the spacecraft would be within 10$^\circ$ of the Sun as viewed from the Earth.  This period of 280 days will occur once every 3 years, provided the proper maneuvers are performed.  Two similar periodic curves give the Sun-Earth-Probe angles for the two spacecraft while the lower smooth curve gives their angular separation as seen from the Earth.

An orbit with a 3:2 resonance with the Earth uniquely satisfy the LATOR orbital requirements\cite{lator_cqg_2004,lator_LCDF-2006}.  For this orbit, 13 months after the launch, the spacecraft are within $\sim10^\circ$ of the Sun with first occultation occurring 15 months after launch\cite{lator_cqg_2004}.  At this point, LATOR is orbiting at a slower speed than the Earth, but as LATOR approaches its perihelion, its motion in the sky begins to reverse and the spacecraft is again occulted by the Sun 18 months after launch.  As the craft slow down and move out toward aphelion, their motion in the sky reverses again, and it is occulted by the Sun for the third and final time 21 months after launch.

%**************
%
%%**************
%\vskip 5pt 

\subsection{Observing Sequence}
\label{sec:operations}

As a baseline design for the LATOR orbit, both spacecraft will be launched on the same launch vehicle. Almost immediately after the launch there will be a 30 m/s maneuver that separates the two spacecraft on their 3:2 Earth resonant orbits (see Figure \ref{fig:lator_sep_angle2}).  This sequence will be initiated at the beginning of the experiment period, after ISS emergence from the Earth's shadow (see Figure~\ref{fig:iss_config}). It assumes that boresighting of the spacecraft attitude with the spacecraft transmitters and receivers have already been accomplished. This sequence of operations is focused on establishing the ISS to spacecraft link. The interspacecraft link is assumed to be continuously established after final deployment (at $\sim15^\circ$ off the Sun), since the spacecraft never lose line of sight with one another. 

The laser beacon transmitter at the ISS is expanded to have a beam divergence of 30 arcsec in order to guarantee illumination of the LATOR spacecraft. After re-emerging from the Earth's shadow this beam is transmitted to the craft and reaches them in about 18 minutes. At this point, the LATOR spacecraft acquire the expanded laser beacon signal. In this mode, a signal-to-noise ratio (SNR) of 4 can be achieved with 30 seconds of integration. With attitude knowledge of 10 arcsec and an array field of view of 30 arcsec no spiral search is necessary. Upon signal acquisition, the receiver mirror on the spacecraft will center the signal and use only the center quad array for pointing control. Transition from acquisition to tracking should take about 1 minute. Because of the weak uplink intensity, at this point, tracking of the ISS station is done at a very low bandwidth. The pointing information is fed-forward to the spacecraft transmitter pointing system and the transmitter is turned on. The signal is then re-transmitted down to the ISS.

Each interferometer station and laser beacon station searches for the spacecraft laser signal. The return is heterodyned by using an expanded bandwidth of 300~MHz. In this case, the solar background is the dominant source of noise, and an SNR of 5 is achieved with 1 second integration\cite{lator_LCDF-2006}. Because of the small field of view of the array, a spiral search will take 30 seconds to cover a 30 arcsec field. Upon acquisition, the signal will be centered on the quad cell portion of the array and the local oscillator frequency locked to the spacecraft signal. The frequency band will then be narrowed to 5 kHz. In this regime, the solar background is no longer the dominant noise source and an SNR of 17.6 can be achieved in only 10 msec of integration. This will allow one to have a closed loop pointing bandwidth of greater than 100 Hz and be able to compensate for the tilt errors introduced by the atmosphere. The laser beacon transmitter will then narrow its beam to be diffraction limited ($\sim$1 arcsec) and to point toward the LATOR spacecraft. This completes the signal acquisition phase, and the entire architecture is in-lock and transmits scientific signal.  This procedure is re-established during each 92-minute orbit of the ISS (see Ref.~\refcite{lator_LCDF-2006} for details).

%We shall now consider the basic elements of the LATOR optical design. 

\subsection{Principles of Optical Design}
\label{sec:optical_design}

\begin{figure*}[t!]
 \begin{center}
\noindent    
\epsfig{figure=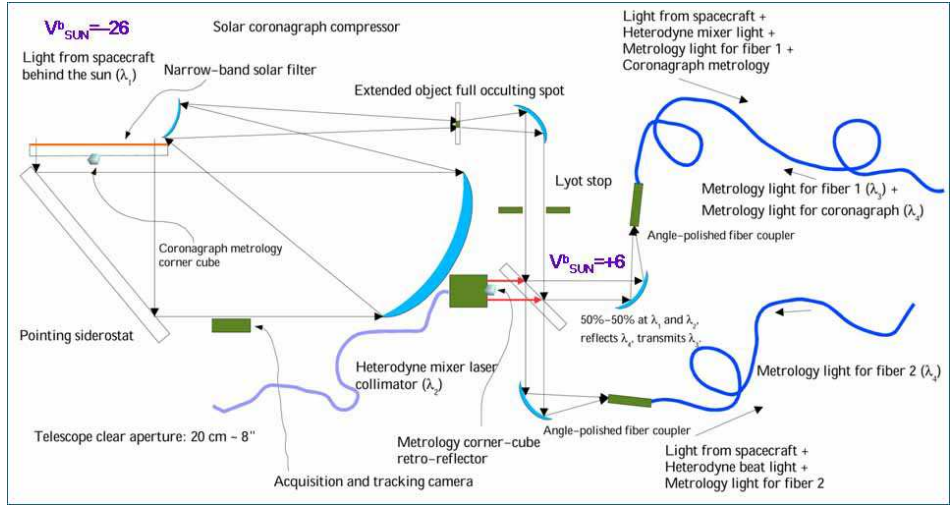,width=125mm}%,height=90mm}
\end{center}
\vskip -10pt 
  \caption{Basic elements of optical design for the LATOR interferometer: The laser light (together with the solar background) is going through a full aperture ($\sim20$cm) narrow band-pass filter with $\sim 10^{-4}$ suppression properties. The remaining light illuminates the baseline metrology corner cube and falls onto a steering flat mirror where it is reflected to an off-axis telescope with no central obscuration (needed for metrology). It then enters the solar coronograph compressor by first going through a 1/2 plane focal plane occulter and then coming to a Lyot stop. At the Lyot stop, the background solar light is reduced by a factor of $10^{6}$. The combination of a narrow band-pass filter and coronograph enables the solar luminosity reduction from $V=-26$ to $V=4$ (as measured at the ISS), thus enabling the LATOR precision observations.
\label{fig:optical_design}}
\vskip -5pt 
\end{figure*}

A single aperture of the interferometer on the ISS consists of three 20 cm diameter telescopes (see Figure \ref{fig:optical_design} for a conceptual design). One of the telescopes with a very narrow bandwidth laser line filter in front and with an InGaAs camera at its focal plane, sensitive to the 1064 nm laser light, serves as the acquisition telescope to locate the spacecraft near the Sun.
The second telescope emits the directing beacon to the spacecraft. Both spacecraft are served out of one telescope by a pair of piezo-controlled mirrors placed on the focal plane. The properly collimated laser light ($\sim$10~W) is injected into the telescope focal plane and deflected in the right direction by the piezo-actuated mirrors. 

The third telescope is the laser light tracking interferometer input aperture, which can track both spacecraft at the same time. To eliminate beam walk on the critical elements of this telescope, two piezo-electric X-Y-Z stages are used to move two single-mode fiber tips on a spherical surface while maintaining focus and beam position on the fibers and other optics. Dithering at a few Hz is used to make the alignment to the fibers and the subsequent tracking of the two spacecraft completely automatic. The interferometric tracking telescopes are coupled together by a network of single-mode fibers whose relative length changes are measured internally by a heterodyne metrology system to an accuracy of less than 5~pm.

%************
%**************

The spacecraft are identical in construction and contain a 1~W, stable (2~MHz per hour $\sim$500 Hz per second), small cavity fiber-amplified laser at 1064~nm. Nearly 75~\% of the power of this laser is pointed to the Earth through a 15~cm aperture telescope and its phase is tracked by the interferometer. With the available power and the beam divergence, there are enough photons to track the slowly drifting phase of the laser light. The remaining part of the laser power is diverted to another telescope, which points toward the other spacecraft. In addition to the two transmitting telescopes, each spacecraft has two receiving telescopes.  The receiving telescope, which points toward the area near the Sun, has laser line filters and a simple knife-edge coronagraph to suppress the Sun's light to 1 part in $10^4$ of the light level of the light received from the space station. The receiving telescope that points to the other spacecraft is free of the Sun light filter and the coronagraph.

The spacecraft also carry a small (2.5~cm) telescope with a CCD camera. This telescope is used to initially point the spacecraft directly toward the Sun so that their signal may be seen at the space station. One more of these small telescopes may also be installed at right angles to the first one, to determine the spacecraft attitude, using known, bright stars. The receiving telescope looking toward the other spacecraft may be used for this purpose part of the time, reducing hardware complexity. Star trackers with this construction were demonstrated many years ago and they are readily available. A small RF transponder with an omni-directional antenna is also included in the instrument package to track the spacecraft while they are on their way to assume the orbital position needed for the experiment. 

The LATOR experiment has a number of advantages over techniques that use radio waves to measure gravitational light deflection. Advances in optical communications technology, allow low bandwidth telecommunications with the LATOR spacecraft without having to deploy high gain radio antennae needed to communicate through the solar corona. The use of the monochromatic light enables the observation of the spacecraft almost at the limb of the Sun, as seen from the ISS. The use of narrowband filters, coronagraph optics and heterodyne detection will suppress background light to a level where the solar background is no longer the dominant noise source. In addition, the short wavelength allows much more efficient links with smaller apertures, thereby eliminating the need for a deployable antenna. Finally, the use of the ISS will allow conducting the test above the Earth's atmosphere---the major source of astrometric noise for any ground based interferometer. This fact justifies LATOR as a space mission.

%*****************************
\section{Conclusions}  
\label{sec:conc}

The LATOR mission aims to carry out a test of the curvature of the solar system's gravity  field with an accuracy better than 1 part in 10$^{9}$. 
The LATOR experiment benefits from a number of advantages over techniques that use radio waves to study the light propagation in the solar vicinity.  The use of monochromatic light enables the observation of the spacecraft almost at the limb of the Sun.  The use of narrowband filters, coronagraph optics, and heterodyne detection will suppress background light to a level where the solar background is no longer the dominant noise source.  The short wavelength allows much more efficient links with smaller apertures, thereby eliminating the need for a deployable antenna.  Advances in optical communications technology allow low bandwidth telecommunications with the LATOR spacecraft without having to deploy high gain radio antennae needed to communicate through the solar corona.  Finally, the use of the ISS not only makes the test affordable, but also allows conducting the experiment above the Earth's atmosphere---the major source of astrometric noise for any ground based interferometer.  This fact justifies the placement of LATOR's interferometer node in space. 

LATOR is envisaged as a partnership between NASA and ESA wherein both partners are essentially equal contributors, while focusing on different mission elements: NASA provides the deep space mission components and interferometer design, while building and servicing infrastructure on the ISS is an ESA contribution\cite{lator_LCDF-2006,ESLAB2005_LATOR}. The NASA focus is on mission management, system engineering, software management, integration (both of the payload and the mission), the launch vehicle for the deep space component, and operations. The European focus is on interferometer components, the initial payload integration, optical assemblies and testing of the optics in a realistic ISS environment. The proposed arrangement would provide clean interfaces between familiar mission elements.

\section*{Acknowledgements}

The work described here was carried out at the Jet Propulsion Laboratory, California Institute of Technology, under a contract with the National Aeronautics and Space Administration.

%%**************************************

%********************************
\end{document}